\documentclass{article}
\usepackage{arxiv}
\usepackage[english]{babel}
\usepackage[utf8]{inputenc} 
\usepackage[T1]{fontenc}    
\usepackage{hyperref}       
\usepackage{url}            
\usepackage{booktabs}       
\usepackage{amsfonts}       
\usepackage{nicefrac}       
\usepackage{microtype}      
\usepackage{lipsum}
\usepackage{multirow}
\usepackage{comment}
\usepackage{cite}
\usepackage{graphicx}
\usepackage{academicons}
\usepackage{xcolor}
\usepackage{scalerel}
\usepackage{tikz}
\usetikzlibrary{svg.path}
\newcommand{\STAB}[1]{\begin{tabular}{@{}c@{}}#1\end{tabular}}
\definecolor{orcidlogocol}{HTML}{A6CE39}
\tikzset{
  orcidlogo/.pic={
    \fill[orcidlogocol] svg{M256,128c0,70.7-57.3,128-128,128C57.3,256,0,198.7,0,128C0,57.3,57.3,0,128,0C198.7,0,256,57.3,256,128z};
    \fill[white] svg{M86.3,186.2H70.9V79.1h15.4v48.4V186.2z}
                 svg{M108.9,79.1h41.6c39.6,0,57,28.3,57,53.6c0,27.5-21.5,53.6-56.8,53.6h-41.8V79.1z M124.3,172.4h24.5c34.9,0,42.9-26.5,42.9-39.7c0-21.5-13.7-39.7-43.7-39.7h-23.7V172.4z}
                 svg{M88.7,56.8c0,5.5-4.5,10.1-10.1,10.1c-5.6,0-10.1-4.6-10.1-10.1c0-5.6,4.5-10.1,10.1-10.1C84.2,46.7,88.7,51.3,88.7,56.8z};
  }
}

\newcommand\orcidicon[1]{\href{https://orcid.org/#1}{\mbox{\scalerel*{
\begin{tikzpicture}[yscale=-1,transform shape]
\pic{orcidlogo};
\end{tikzpicture}
}{|}}}}

\title{Young Adult Unemployment Through the Lens of Social Media: Italy as a case study.}

\author{
  Alessandra Urbinati\orcidicon{0000-0002-9031-1763}\thanks{These authors contributed equally to this work. Please address any correspondence to \href{mailto:alessandra.urbinati@unito.it}{alessandra.urbinati@unito.it} and \href{mailto:kyriaki.kalimeri@isi.it}{kyriaki.kalimeri@isi.it}} \\
  Department of Computer Science\\
  University of Turin\\
   \And
   Kyriaki Kalimeri \orcidicon{0000-0001-8068-5916}\\
   ISI Foundation,  Turin,  Italy\\
   \And
   Andrea Bonanomi\orcidicon{0000-0003-2857-1430}\\
   Universit\`{a} Cattolica del Sacro Cuore (UNICATT), Milan, Italy\\
   \And
   Alessandro Rosina\orcidicon{0000-0002-0158-0583}\\
   Universit\`{a} Cattolica del Sacro Cuore (UNICATT), Milan, Italy\\
   \And
   Ciro Cattuto\orcidicon{0000-0001-9526-4364} \\
   ISI Foundation,  Turin,  Italy\\
   \And
Daniela Paolotti\orcidicon{0000-0003-1356-3470}\\
   ISI Foundation,  Turin,  Italy\\
   
}

\begin{document}
\maketitle

\begin{abstract}
Youth unemployment rates are still in alerting levels for many countries, among which Italy. Direct consequences include poverty, social exclusion, and criminal behaviours, while negative impact on the future employability and wage cannot be obscured.
In this study, we employ survey data together with social media data, and in particular likes on Facebook Pages, to analyse personality, moral values, but also cultural elements of the young unemployed population in Italy. Our findings show that there are small but significant differences in personality and moral values, with the unemployed males to be less agreeable while females more open to new experiences. 
At the same time, unemployed have a more collectivist point of view, valuing more in-group loyalty, authority, and purity foundations. 
Interestingly, topic modelling analysis did not reveal major differences in interests and cultural elements of the unemployed. Utilisation patterns emerged though;
the employed seem to use Facebook to connect with local activities, while the unemployed use it mostly as for entertainment purposes and as a source of news, making them susceptible to mis/disinformation.
We believe these findings can help policymakers get a deeper understanding of this population and initiatives that improve both the hard and the soft skills of this fragile population.
\end{abstract}


\section{Introduction}
Youth unemployment is one of the most significant challenges that modern societies face, with more than 3.3 million unemployed young people (aged 15-24 years) in 2019 in the EU alone. 
In 2014, youth unemployment in Italy reached 46\%~\cite{istat}. Today, this rate is at 37\%~\cite{istat}, substantially lower than five years ago, but still in alerting levels and well above the average European trend\footnote{\url{https://ec.europa.eu/social/main.jsp?catId=1036}}.
The ``dejuvenation'' of the Italian population exacerbates this phenomenon~\cite{Caltabiano2018}, urging for a better understanding of this vulnerable segment of the society and devising better policies.

Drivers of youth unemployment vary over place and time: inexperience in connecting with the job market, limited job offers, and lack of qualifications are some of the most common determinants \cite{o2015five,baah2016youth,mayer2018crisis}. 
The social and economical impact of high unemployment rates among the younger age groups cannot be overstated.
High inactivity rates create a vicious circle of poverty, homelessness, social exclusion \cite{gallie2003unemployment}, marginalisation~\cite{Hallsten2017}, and
 criminal behaviours \cite{freeman1994crime,buonanno2006crime}.
At an individual level, youths are in a fragile phase where the hardships of getting a job can cause pronounced, long-term feedback effects leaving a negative impact on later employability \cite{gregg2001impact,Proserpio2016,furnham20152} and the future wage \cite{gregg2005wage}. 
Early unemployment also has a potential impact on physical and mental health \cite{hammer1993unemployment,Strandh2014}, as well as psychological distress, depression, and suicides \cite{hammarstrom2002early,Adler2019}. 

Traditionally, scientists gained insights into this phenomenon via survey administration to representative population groups~\cite{clark1982dynamics}.
In this work, we adopt a mixed approach employing both traditional surveys administered on Facebook, and large scale observational digital data.
We engaged a large cohort in Italy via a Facebook-hosted app, whose main functionality was to administer a series of questionnaires regarding personality traits, moral values, general interests, and political views, while at the same time, gauged the participants' Likes on Facebook Pages.
Observing these digital behaviours, we aim to shed light on facets of youth unemployment that are difficult, or even impossible, to assess via traditional means.

We contribute to the current state of the art by exploring associations between moral values, political views and employability, accounting at the same time for demographic, behavioural and geography bias.
Our findings show that the employed appear to cherish individualistic values. On the other hand, the unemployed tend to trust more the EU and the national government. At the same time, they are less open to the introduction and assimilation of migrants in society, and they are particularly worried about protecting their traditional values.
Thoroughly analysing the digital patterns of interests, as expressed through their Page ``Likes'', we found that the macro interests of both communities largely overlap. However, the way employed and unemployed utilise the Facebook platform presents subtle differences. 

These results push back against conventional ideas about professional success; instead, they show proof of the fact that unemployment is not driven by a lack of interests, rather than a considerable lack of opportunities.
Finally, we believe that these findings can inform policymakers to support initiatives that do not solely improve the hard but also the soft skills of this fragile population.

\section{Background and Related Work}

The role of psychological attributes in the labour market has received increasing interest in economics and organisation psychology during the past years. Extended literature across various scientific communities highlights for instance the relationship between personality traits and employability skills, e.g.~\cite{viinikainen2012personality,finnerty2016stressful,kalimeri2010causal} (see~\cite{almlund2011personality} for a review). 
Getting a job that is in line with one's personality traits was shown to lead to higher career satisfaction~\cite{lounsbury2008personality}, higher productivity~\cite{finnerty2014towards,kalimeri2013going}, and consequently, well-being and health quality~\cite{friedman2014personality}.
However, evidence on the relationship between personality traits and unemployment is still scarce~\cite{uysal2011unemployment,viinikainen2012personality,bonanomi2017understanding}.
Limited studies also exist for other psychological constructs such as human values~\cite{sortheix2015work} and culture~\cite{brugger2009does}, which were recently shown to be related to individuals' ability to connect to the job market.

New data sources such as search engine queries, mobile phone records, and social media data, allow for unobtrusive observation of both psychological and behavioural patterns~\cite{Kalimeri2019predicting,Kosinski2012}.
Pioneering work in predicting unemployment rates from search engine query data~\cite{Ettredge2005} paved the way for other nationally-wide studies which proved the feasibility of observing and predicting unemployment rates from query data~\cite{Gao2019}, also in small and emerging economies~\cite{pavlicek2015nowcasting}.

More recently, social media platforms were exploited to create useful indicators of the socio-economical status of population groups that are very hard to reach with more traditional means of study~\cite{Rama2020}.
Analysing Twitter data~\cite{Antenucci2014} created a social-media based index of job loss, which tracks the initial unemployment insurance claims at medium and high frequencies. 
Llorente et al.~\cite{Liorente2015} with geotagged tweets collected from Spain, found a strong positive correlation between unemployment and the Twitter penetration rate, due to diversity in mobility fluxes, daily rhythms, and grammatical styles of Twitter users. 
Bokanyi et al.~\cite{Bokanyi2017} studied how employment statistics of counties in the US are encoded in the daily rhythm of people. At the same time, Proserpio et al.~\cite{Proserpio2016} explored the relations between psychological well-being and the macroeconomic shock of employment instability, showing that SM can capture and track changes in psychological variables over time.
Employing Facebook data~\cite{bonanomi2017understanding}, predicted youth unemployment, and in particular the NEET status (Not in Employment, Education, or Training population), leveraging on people's  ``Likes'' on Facebook Pages.

In this study, we are not interested in predicting the unemployment status or rate; rather we focus on the association between political views and moral values to employability, which, in contrast to other psychological traits, is largely unexplored.
Additionally, we employ Facebook as a social sensor, focusing on the understanding of interests and cultural elements from the digital patterns of Facebook Page ``Likes''. We provide, in a data-driven approach, the important topics of interest of the employed and unemployed, but also insights on how the communities utilise the Facebook Platform.

\section{Data Collection}
\vspace{-5pt}

The data were collected employing a Facebook-hosted application, designed to act as a surveying tool since a recent study showed that Facebook is a valid scientific tool to administer demographic and psychometric surveys~\cite{Kalimeri2020}.
The app was disseminated through a traditional media campaign. Upon providing their informed consent, participants agreed to provide us with their ``Likes'' on Facebook Pages.
After entering the app, the participants were requested to provide basic demographic information, namely gender, employment status and province of residence. Then, they proposed a series of questionnaires regarding
(i) psychometric attributes, (ii) general cultural interests, and (iii) political opinions. Inserting their demographic information as well as completing the questionnaires and surveys was on a strict volunteering basis.
The application was mainly deployed in Italy; it was initially launched in March 2016, while the data used here were downloaded in September 2019. 

\textbf{Demographic Information.}
Table~\ref{tab:demographic} reports the demographic information of the participants. Data are not complete, given that the form was not mandatory. We report the percentages for gender, age, employment status and location of the general population as obtained from the official national statistics for 2019~\cite{istat} (Census). Next, we report the total of individuals that have accessed the app (Facebook App) and the individuals that have declared their qualification and are under 44 years of age (Complete Dataset). 
The Balanced Dataset consists of a sample of the Complete Dataset aimed to avoid gender, employment, location, as well as digital activity bias. More details are reported in the ``Data processing'' section.

\textbf{Psychometric Surveys.} 
The Big Five Personality Dimensions \cite{gosling2003very} and the Moral Foundations Questionnaire \cite{graham2009liberals,Haidt2007,Haidt2004} were administered. Both are scientifically validated and extensively employed in the literature to assess personality and morality, respectively.
The Big Five Personality Dimensions (BIG5) personality traits model characterises personality based on five dimensions,
(i) Openness to experience,
(ii) Conscientiousness,
(iii) Extraversion,
(iv) Agreeableness,
(v) Neuroticism.
MFT focuses on the psychological basis of morality, identifying the individualistic group consisting of the(i) Care/Harm, and (ii) Fairness/Cheating foundations, and the social binding group consisting of the (iii) Loyalty/Betrayal,
(iv) Authority/Subversion, and 
(v) Purity/Degradation foundations. 

\textbf{Additional Surveys.} 
The app also proposed a series of surveys aimed at addressing broader political topics but also the general opinions of participants.
Inspired by the 41-item inventory proposed by Schwartz et al.~\cite{schwartz1992universals} and the updated version of Barnea et al.~\cite{barnea1998values} adapted to the Italian context, the Political Opinions (PO) survey explores essential societal issues. The survey consists of the following 8-items from the original inventory:
 \textit{Q1} - It is extremely important to defend our traditional religious and moral values, 
  \textit{Q2} - People who come to live here from other countries generally make our country a better place to live,
  \textit{Q3} - I trust the President of the Republic,
  \textit{Q4} - I believe every person in the world should be treated in the same way. I believe that everyone should have the same opportunities in life,
  \textit{Q5} - I trust the national government,
  \textit{Q6} - The less the government gets involved with the business and the economy, the better off this country will be,
  \textit{Q7} - I trust the European Union,
  \textit{Q8} - I strongly believe that the state needs to be always aware of the threads, both internal and external.
In the Culture and Interests (CI) survey, participants were asked to rate their interest in several categories in a 5-point Likert scale: Travel, Sport, Science, Food, Culture, Nature, Society and Politics, Education, Health, Hobbies, Business, Shopping.

\begin{table*}[ht]
    \centering
    \renewcommand{\arraystretch}{1.2}

    \begin{tabular}{llcccc}
        \textbf{}&\textbf{}&\textbf{Census}&\textbf{Facebook}&\textbf{Complete}&\textbf{Balanced}\\
        \textbf{}&\textbf{}&\textbf{}&\textbf{ App}&\textbf{Dataset}&\textbf{Dataset}\\
        & & & \textit{n = 63,980}& \textit{n = 3079}& \textit{n = 842} \\\toprule
        \textbf{Gender}&Female&51.6\%&31.7\%&39.6\%&50\%\\
        &Male&48.4\%&68.3\%&60.4\%&50\%\\\midrule
        \textbf{Age}&16-17&4.8\%&1.6\%&0.1\%&0.1\%\\
        &18-24&4.9\%&38.3\%&12.8\%&16.7\%\\
        &25-34&11.0\%&28.5\%&56.0\%&55.9\%\\
        &35-44&13.8\%&13.0\%&24.2\%&27.2\%\\
        &45+&53.9\%&18.7\%&-&-\\\midrule
        \textbf{Qualification}&Employed&54.2\%&42.4\%&80.2\%&50\%\\ 
        &Unemployed&8.6\%&7.2\%&19.8\%&50\%\\
        &Others(*)&37.2\%&50.4\%&-&-\\
        \midrule
        \textbf{Location}&North&44.1\%&53.7\%&50.8\%&39.4\%\\
        &Center&19.5\%&18.5\%&19.5\%&26.5\%\\
        &South&36.4\%&29.2\%&29.2\%&34.1\%\\
        \bottomrule
    \end{tabular}
    \caption{Demographic breakdown of our data according to gender, age, qualification, and geographic location. The first column, ``Census'', reports the  percentages per attribute according to the statistics provided by the official census bureau~\cite{istat}. 
    The ``Facebook App'' column reports the percentages of the total number of participants. We use ``Others'', in the qualification section, as an umbrella term that includes students, or missing responses. 
    The ``Complete Dataset'' consists only of the people that have declared their qualification and are under 44 years of age. 
    Finally, the ``Balanced Dataset'' consists of a random sample of the ``Complete Dataset'' taking into consideration gender, qualification, and location biases. 
    \vspace{-20pt}
\label{tab:demographic}}
\end{table*}

\subsubsection{Facebook Data.}
For the participants who entered the app, we gathered information on the public Facebook Pages which they ``liked'', as well as some necessary metadata about the page, such as the name,  description,  category\footnote{Facebook Pages, are assigned to one of the predefined Facebook categories. Link to the full list of Facebook Categories: \url{https://www.facebook.com/pages/category/}.}, and the popularity of the page in terms of followers. Importantly, all the linguistic information available to us is the content by the page creator. No information about the posts or the comments of the page is available for this analysis. Table~\ref{tab:user_activity} reports the total number of Facebook Pages, Likes, and categories collected per qualification status.

\begin{table}[ht]
    \centering
    \renewcommand{\arraystretch}{1.2}
    \begin{tabular}{lcccc}
    &&\textbf{Complete Dataset} & \textbf{Balanced Dataset} & \\
    &&\textit{n = 3079} & \textit{n = 842} &\\
    \toprule
    \textbf{$\#$Pages}&Employed&730,237&189,199  (just E 41,151)\\
    &Unemployed&307,916&229,548 (just U 58,275)\\
    \midrule
    \textbf{$\#$Likes}&Employed& 2,063,944&332,610\\
    &Unemployed&780,693&456,971\\
    \midrule
    \textbf{$\#$Category} & &1,396 & 1,214\\
    \bottomrule
    \end{tabular}
    \caption{Total number of unique Facebook Pages and total number of Likes per qualification type in the ``Complete Dataset'' and the ``Balanced Dataset''. We report the number of pages liked only by the employed or unemployed, respectively. We also report the total number of unique Facebook categories in the two datasets. 
\label{tab:user_activity}}
\end{table}

\subsection{Data processing}

\begin{figure}[ht]
\centering
\includegraphics[width=0.6\columnwidth]{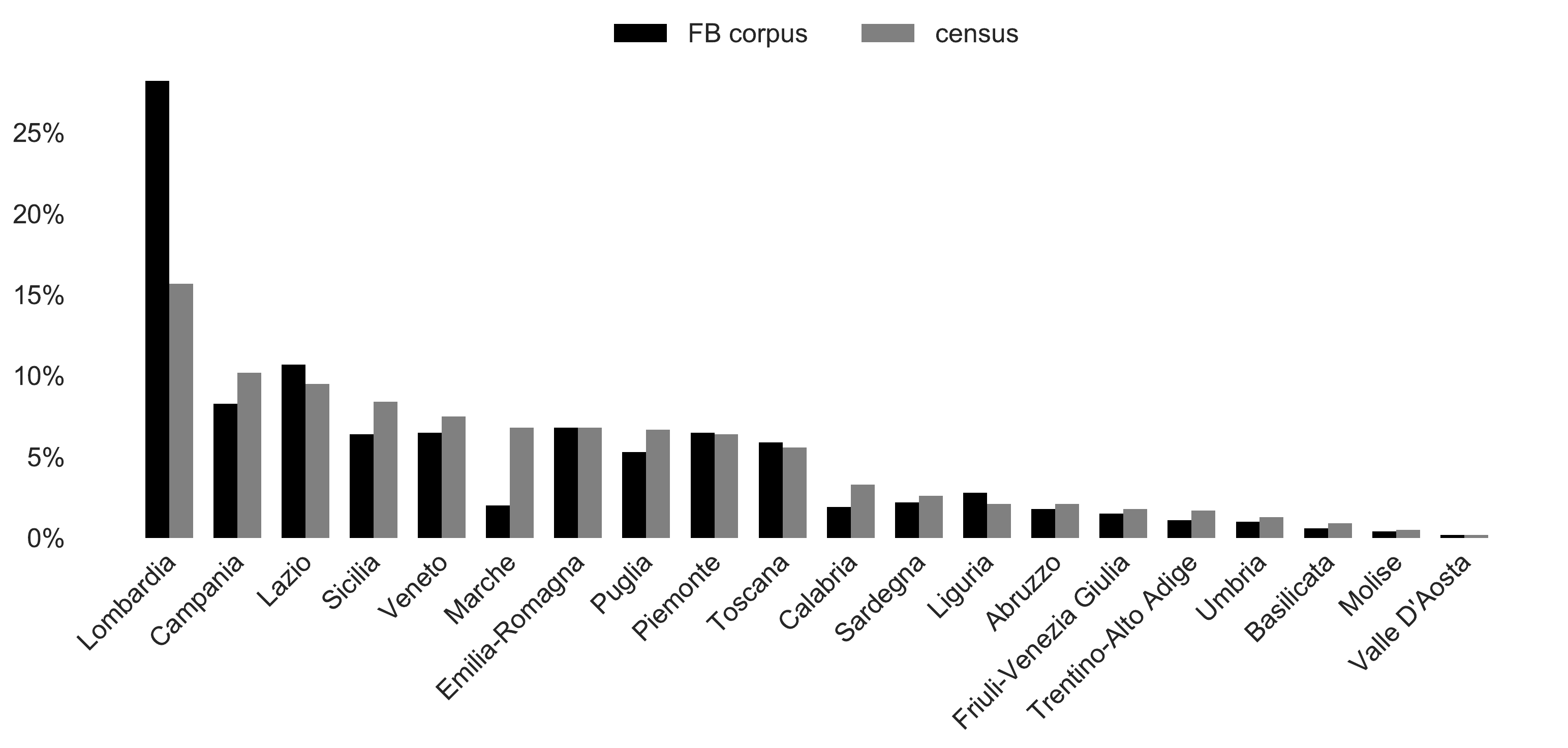}
\vspace{-10pt}
\caption{Geographic distribution of the population in the Complete Dataset (dark-coloured bars) as compared to the expected population distribution according to the official census data per region (light-coloured bars).}
\label{fig:pop_chis}
\vspace{-10pt}
\end{figure}

\textbf{Demographic Information.} From the initial cohort of the app, we derived the ``Complete Dataset'', a subset of individuals for which we have complete information records about their gender, employment status, and geographic location, and who are younger than 44 years old.
Table~\ref{tab:demographic} reports the demographic breakdown of the population according to the official census as well as the datasets employed in this study.
Lombardy region is over-represented, a phenomenon explained by the fact that the project was initially launched in that region. At the same time, we notice that the area of Marche is under-represented, while all other regions in our cohort follow the distribution of the census closely\footnote{Figure~\ref{fig:pop_chis} in SI provides a clear picture of the geographical distribution per region as compared to the expected values with respect to the census.}.
Our cohort deviates from the census distribution for gender, with males to be over-represented. We also have an over-representation of the 18-24 and 25-34 age groups.
In terms of qualification, we notice that we have an over-representation of employed individuals.

To avoid gender and qualification biases (see Table~\ref{tab:demographic}) which affect the linguistic analysis and the topic modelling, we created a subset of the Complete Dataset, namely  ``Balanced Dataset''. The Balanced Dataset consists of randomly selected participants who are equally distributed for gender, qualification, and geography.
To obtain this subset, firstly, we identified $N_{min}$, the population of the smallest demographic subgroup in our dataset, in terms of gender, qualification, and location breakdown. 
Then, we divided our participants into four quadrants according to their ''likes per user'' activity. From each quadrant, we randomly sampled $N_{min}$ participants, resulting with a total population that consists of $4 \star N_{min}$ participants uniformly distributed with respect to age, gender, and geography.
The ``Balanced Dataset'' is much smaller, however, less impacted by demographic and activity biases. It is employed only in the language and topic modelling analysis of the following section.

\textbf{Facebook Data.} Among the Pages liked by participants in the sample we retain for the analysis, 65\% of them are in Italian, 28\% in English, 2\% in French, and 5\% in other languages. 
For this study, we analyse only the pages in Italian and English. We dropped the generic Page category ``community'', since it may contain pages with diverse content, and also the page of the Italian newspaper ``Repubblica'' due to a communication campaign carried out on that journal.

\vspace{-5pt}
\section{Methods and Results}
\vspace{-5pt}

\subsubsection{Questionnaire Analyses.}
\label{sec:quest}

We assessed the relationship of unemployment to the demographic and psychological attributes by means of logistic regression. A strong statistical significant effect of gender emerged (results reported in Table~\ref{tab:regression} in SI), with women to be more often jobless as confirmed by the literature~\cite{azmat2006gender}. Geography too emerges as a significant factor, with the southern Italian regions to be more heavily affected as supported by the literature~\cite{manacorda2006regional}.

To avoid the known gender biases of the psychological attributes~\cite{feingold1994gender,tangney2002gender}, we divided the analysis per gender.
We assess differences in the psychological attributes, interests, and political views between the employed (E) and unemployed (U) populations using the Mann-Whitney U test ~\cite{nachar2008mann}, since our data are ordinal and not necessarily follow a normal distribution. 

Personality Traits. 
We assess differences in the personality traits between the employed and unemployed populations, employing the Mann-Whitney U test. As reported in Table~\ref{tab:quest}, employed males are more agreeable while unemployed females are more open to new experiences. 

Moral Foundations.
Unemployed females value significantly more the protection of others, they value more the in-group loyalty, and respect to authority. Both males and females have a higher purity score, and hence exhibiting a more social binding worldview (see Table~\ref{tab:quest}).
The relationship of moral values with unemployment has received much less attention by the scientific community, despite playing a critical motivational role in job selection and successful school-to-work transition~\cite{sortheix2015work,sortheix2013role}, and the disengagement from the labour market~\cite{barr2016moral}.

Political Opinions.
Turning to the political views of the two populations, we find that unemployed males tend to trust less the EU as well as the national government and president. Along the same lines, female unemployed trust less the presidency and national government, while they see the privatisation of the public enterprises and the minimisation of the governmental involvement in the economy~\cite{westholm1986youth} positively. 
Employed males feel less the need to protect their traditional values and believe in the fair treatment of individuals believe the state should always be aware of national and international threats.
The current literature~\cite{finseraas2016going} also supports several of these such findings reported in Table~\ref{tab:quest}.

\begin{table*}[h]
    \centering
        \renewcommand{\arraystretch}{1.1}
\begin{tabular}{lccc|ccc}
\textbf{Big Five} &  \textbf{EM} & \textbf{UM} & \textbf{pvalue} & \textbf{EF} & \textbf{UF} &\textbf{pvalue}  \\
\toprule
Extraversion &   8.83$\pm$0.21 &   9.02$\pm$0.52 &          - &   8.94$\pm$0.17 &   9.04$\pm$0.26 &          - \\
     Agreeableness &   9.41$\pm$0.24 &   8.83$\pm$0.54 &          * &   9.55$\pm$0.21 &   9.66$\pm$0.26 &          - \\
 Conscientiousness &   9.16$\pm$0.22 &   9.47$\pm$0.45 &          - &   9.27$\pm$0.19 &   9.61$\pm$0.27 &          - \\
          Openness &   8.74$\pm$0.24 &   8.98$\pm$0.52 &          - &   9.26$\pm$0.22 &   9.59$\pm$0.33 &          * \\
      Neurotisism &   9.91$\pm$0.22 &  10.02$\pm$0.51 &          - &   9.93$\pm$0.20 &  10.20$\pm$0.27 &          - \\
\textbf{Moral Foundation} & & & &  & &  \\
\toprule
Care &  18.18$\pm$0.34 &  18.32$\pm$0.81 &          - &  18.77$\pm$0.33 &  19.21$\pm$0.45 &          * \\
           Loyalty &  15.33$\pm$0.40 &  15.96$\pm$0.89 &          - &  15.94$\pm$0.41 &  16.82$\pm$0.52 &        *** \\
         Authority &  14.48$\pm$0.42 &  14.84$\pm$0.96 &          - &  15.43$\pm$0.39 &  16.32$\pm$0.54 &        *** \\
            Purity &  14.56$\pm$0.46 &  15.84$\pm$1.05 &         ** &  16.40$\pm$0.41 &  17.33$\pm$0.57 &         ** \\
     Individualism &  38.64$\pm$0.54 &  38.01$\pm$1.49 &          - &  39.13$\pm$0.57 &  39.63$\pm$0.82 &          * \\
     SocialBinding &  44.37$\pm$1.09 &  46.65$\pm$2.56 &          * &  47.77$\pm$1.07 &  50.47$\pm$1.44 &        *** \\
\textbf{Political Opinion} & & & &  & &  \\
\toprule
         Q1 &   2.89$\pm$0.07 &   3.13$\pm$0.19 &         ** &   2.90$\pm$0.13 &   2.97$\pm$0.23 &          - \\
                Q2 &   3.27$\pm$0.06 &   3.03$\pm$0.16 &         ** &   3.34$\pm$0.11 &   3.19$\pm$0.25 &          - \\
                Q3 &   3.26$\pm$0.07 &   2.90$\pm$0.18 &       **** &   3.30$\pm$0.12 &   3.03$\pm$0.25 &          * \\
                Q4 &   4.43$\pm$0.05 &   4.27$\pm$0.14 &          * &   4.56$\pm$0.08 &   4.43$\pm$0.18 &          - \\
                Q5 &   2.74$\pm$0.06 &   2.51$\pm$0.17 &         ** &   2.73$\pm$0.12 &   2.46$\pm$0.24 &          * \\
                Q6 &   2.81$\pm$0.06 &   2.86$\pm$0.17 &          - &   2.77$\pm$0.12 &   2.98$\pm$0.21 &          * \\
               Q7 &   3.47$\pm$0.07 &   3.28$\pm$0.19 &          * &   3.66$\pm$0.12 &   3.42$\pm$0.27 &          - \\
               Q8 &   3.79$\pm$0.06 &   3.96$\pm$0.15 &         ** &   3.76$\pm$0.11 &   3.69$\pm$0.24 &          - \\

\bottomrule
\end{tabular}
    \caption{Mann-Whitney U test between the employed and unemployed populations per gender. Averages and confidence intervals of the comparison between the Male Employed (ME) versus Male Unemployed (MU) and the Female Employed (FE) versus the Female Unemployed (FU) with the respective p-values. Only statistical significant findings are reported here. Please check the SI section for the complete Table. Statistical significance is reported as follows: p-value - "****": $p < 0.0001$, "***": $p < 0.001$, "**": $p < 0.01$, "*": $p < 0.05$, otherwise $-$ for non-significant results.}
    \label{tab:quest}
    
\end{table*}

\begin{figure*}[ht]
  \centering
  \vspace{-10pt}
  \includegraphics[width=1\linewidth]{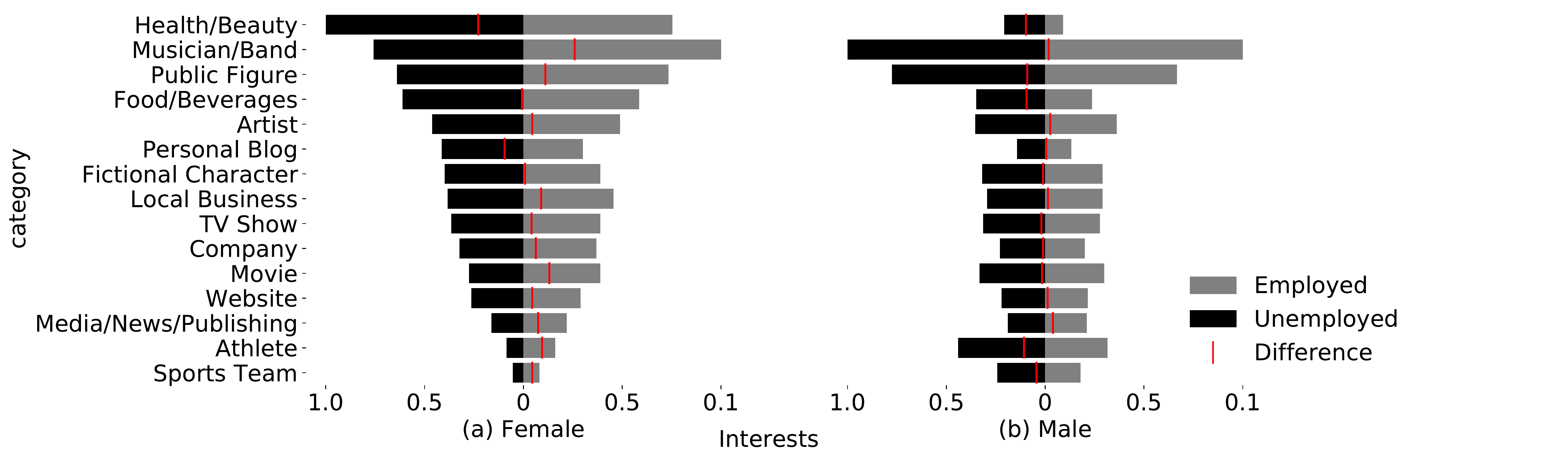}
  \vspace{-20pt}
  \caption{Top 15 Facebook Page categories liked by the Employed (light-coloured bars) and Unemployed (dark-coloured bars) participants in the Complete Dataset. 
  The category frequencies are normalised per gender and qualification.
  The red vertical lines indicate the direction of the difference between the two populations.   
  }
  \label{fig:likes_cat}
\end{figure*}

Culture and Interests.
As reported in Table~\ref{tab:demographic}, in the Complete Dataset, we have very different representation in the participant's demographics. To overcome this issue, we normalised the category frequencies for each gender and qualification subgroup. 
Merging the ten most frequently liked categories by each subgroup, i.e. unemployed female, employed male and so on, we obtained the 15 most frequently liked categories by females (a) and males (b).
Figure~\ref{fig:likes_cat} depicts this distribution.
Marked differences are observed among the female population, with unemployed women to be more interested in health and beauty pages and personal blogs to the employed ones. After visual inspection, these blogs refer to beauty advice and recipes. Their employed counterparts are slightly more interested in musicians and bands, movies, news magazines, and public figures like athletes.
In the male cohort, there are less pronounced differences. Unemployed are interested in health and beauty, public figures, athletes, food and beverages, while employed are more fond of music bands, artists, and news magazines.
Health and beauty pages are the only common interest both for male and female unemployed, probably since these Pages often promote discounts.

\begin{figure*}[ht]
  \centering
    \vspace{-10pt}
  \includegraphics[width=\textwidth]{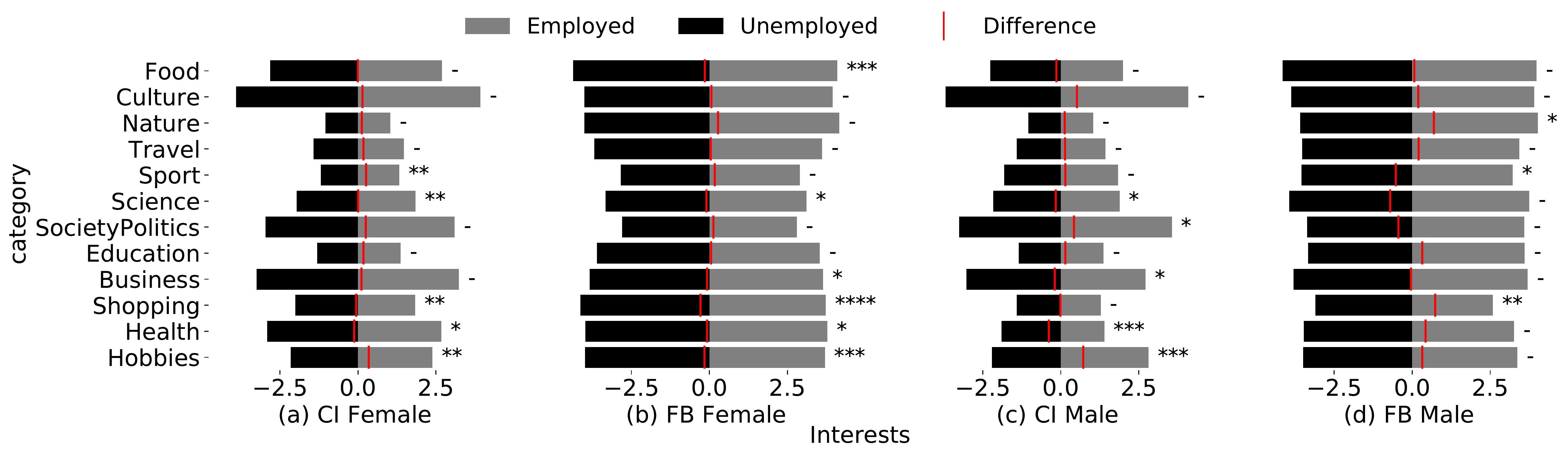}
   \vspace{-15pt}
  \caption{Mann-Whitney U test between the actual (FB) and the self-reported interests (CI) by the Employed (light-coloured bars) and the Unemployed (dark-coloured bars) participants. 
  The Pages categories are manually classified in the twelve macro-categories of the CI questionnaire.
  The bar amplitude corresponds to the median values for the Employed and Unemployed respectively, while the red vertical lines indicate the direction of difference between the two means. 
  Statistical significance is reported as follows: p-value - "****": $p < 0.0001$, "***": $p < 0.001$, "**": $p < 0.01$, "*": $p < 0.05$, otherwise $-$ for non-significant results.}
  \label{fig:likes_quest}
  \vspace{-10pt}
\end{figure*}

Mann-Whitney U test on the CI survey's self-assessments, shows that the employed population declares to be more interested in topics that regard the society and politics and hobbies ((see Figure~\ref{fig:likes_quest} (a) and (c)). 
Mapping the original page categories (1,396 in total) to the twelve more generic interest categories of the CI survey, we assessed the extent to which the self-reporting patterns of interests correspond to the actual digital ones.
The mapping is primarily based on Facebook's original hierarchical structure, where for instance, the various medical specialities are under the generic term ``Medical and Health''. Then, we manually classified a limited amount of categories that were not part of the twelve generic categories of the questionnaire, for instance, ``Vitamins and Supplements'' under ``Health'', ``Non-Profit Organization'' under ``Society'', ``Elementary School'' under ``Education''.
An obvious limitation of this approach is that the frequency of visit to the Liked Page is not available to us, still, liking a page is a clear signal of interest.
Importantly, and despite the limitations, Figure~\ref{fig:likes_quest} (b,d) shows that the responses in the online questionnaires reflect the actual online activity of the users, with only exceptions shopping and hobbies that emerge to be significantly more of interest to the female unemployed population, and health and sports for the unemployed male population.

\subsection{Language and Topic Modelling}

Often, category information is too generic; for instance, a business page could refer to various types of business, like restaurants, or beauty centres. 
To overcome this issue, we focus on the linguistic content of pages' metadata, which includes the page about, and the page description. 
To avoid demographic and behavioural biases, we analysed the Pages included in the ``Balanced Dataset''(see Table~\ref{tab:demographic} and~\ref{tab:user_activity}).
We employed statistical language modelling to distinguish the two communities.  
Considering the differences between the probability distribution of words occurring in pages liked by the two populations and the overall probability distribution of words in the entire dataset, we can identify the most distinguishing words. 
Here, we performed a systematic bi-gram analysis which surfaces common sub-sentences that tend to occur frequently in the text. 
Table~\ref{tab:distinguishingwords} reports the 20 most distinguishing words in descending order.
Pages liked by the unemployed are talking about hobbies; fashion, photography and YouTube channels; on the contrary, the employed ones are about the territory, and projects promoting the local food and culture.
This snapshot of the two communities leads us to think that the employed and unemployed population use the platform differently. 
The unemployed like Facebook Pages to keep updated with their hobbies in a relaxed and fun way but also as a news source, while the employed use the platform to stay in touch with the local communities. 

\begin{table}[ht]

\centering
\begin{tabular}{c c p{0.8\columnwidth}}
\toprule
\multirow{3}{*}[-0.9ex]{\STAB{\rotatebox[origin=c]{90}{\textbf{Unemployed}}}} &IT & Roma, series, film, INTER, welcome, Napoli, love, fashion, creation, advertisement, football, direction, good, photo, passion, best, channel, jewellery, director, beauty\\
&EN & music, fashion, love, film, game, time, series, band, news, years, album, team, products, released, house, games, Italy, life, based, people\\
[0.2cm]\cmidrule{3-3}
\multirow{3}{*}[-3.1ex]{\STAB{\rotatebox[origin=c]{90}{\textbf{Employed}}}}
&IT & association, activity, project, territory, Calabria, culture, cultural, Salerno, centre, Bologna, Genova, cuisine, restaurant, municipal, Reggio, wine, local, social, trapped, research\\
&EN & nail, polish, health, nutrition, architecture, design,
Salento, BMW, food, healthy, diet, Versailles,
steampunk, cancer, science, gothic, lacquer, SAT, disease, medical\\
\multirow{3}{*}[-3.1ex]{\STAB{\rotatebox[origin=c]{90}{\textbf{}}}}
\\
\bottomrule
\end{tabular}
\caption{Top 20 distinguishing words for the employed and unemployed communities for pages in Italian(IT) and English(EN), respectively.}
\label{tab:distinguishingwords}

\end{table}

\begin{figure*}[ht]
\centering
\vspace{-10pt}
\includegraphics[width=\columnwidth,keepaspectratio]{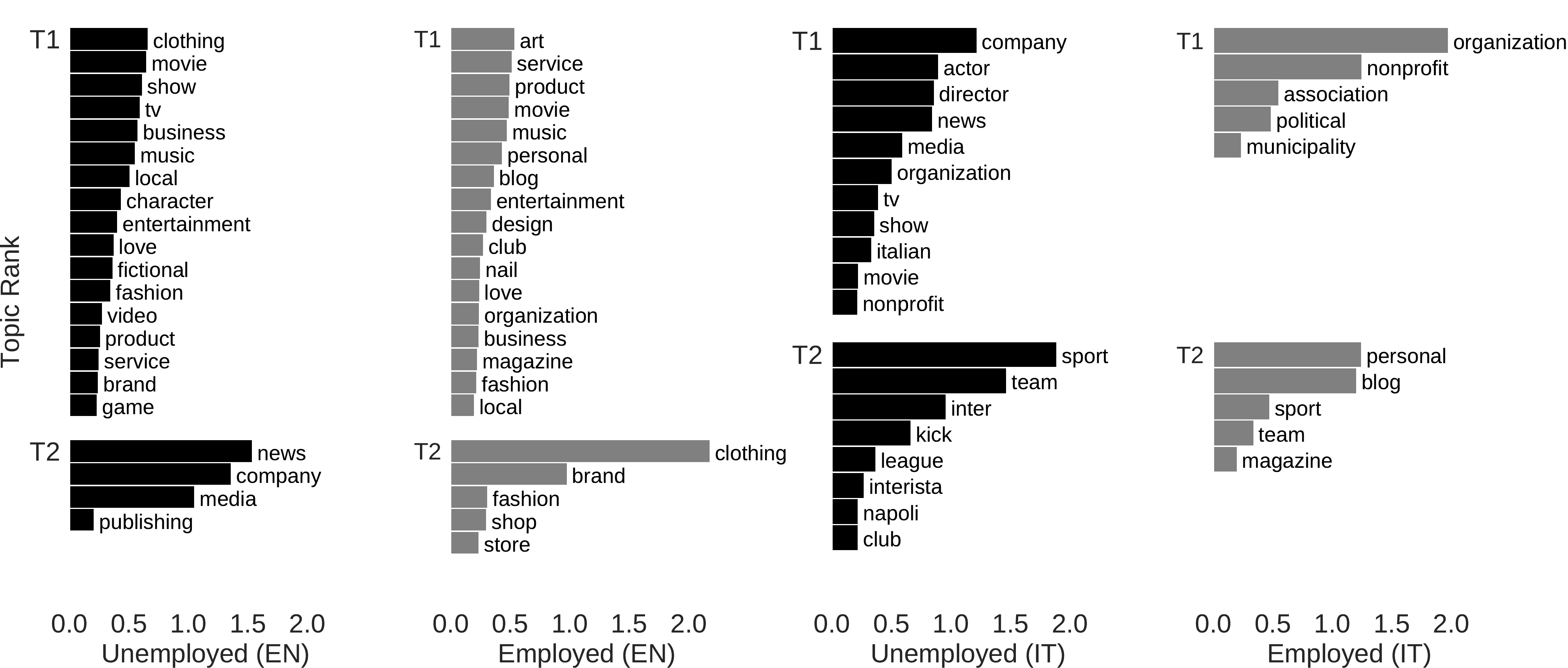}
\vspace{-10pt}

\caption{\label{fig:words_bar}The ten topics according to the NMF algorithm with the most contributing words per topic, for the Italian and English language, respectively. See the Appendix for the complete topic presentation.}
\end{figure*}

Moving to higher level constructs, we aim to uncover latent behavioural patterns in the interest and culture of the two populations.
We apply a topic modelling approach based on the NMF algorithm~\cite{cichocki2006new}, to analyse the linguistic information contained in the Page title, about information, and the full description. We included all pages liked by the population of interest, excluding ones liked by both groups (see Table~\ref{tab:user_activity}).
We create a model per language and qualification for $k=10$ components. 
Figures~\ref{fig:words_bar} depict the most important words; both the topics and the words within each topic are raked in decreasing weight order\footnote{\label{foot:space}Due to space limitations the full topic descriptions are presented in the Appendix, please refer to Figures~\ref{fig:topic_italian} for the Italian and English languages respectively.}.

Noteworthy is the fact that the communities exhibit several differences in the first two topics (T1 and T2) while from the third topic (T3) onward, the differences are negligible. 
Interestingly, the primary topic of the unemployed community regards entertainment TV shows and news, while the respective one for the employed is about non-profit organisations. In contrast, the news appears only in the fourth topic.  
This is of significant importance since it indicates that the unemployed use the Facebook platform to get news information.
Observing the emergent topics, we notice that there are no striking differences between the two communities.  
Another important note lies in the fact that our cohort consists of its vast majority by Italians. This influences the content differences between the topics in Italian and English language; in English, prevalent topics are entertainment, art, and TV shows. Interestingly, both employed and unemployed exhibit the same behavioural pattern.

\vspace{-10pt}

\section{Discussion and Conclusions}\vspace{-5pt}
Youth unemployment in Italy is still in alerting levels, with numerous implications both for the individual and the society.
Engaging a large cohort of young Italians on the Facebook Platform, we provide a more in-depth understanding of psychological, cultural, but also differences in digital behaviours between the employed and unemployed communities.

\textit{Are there associations between psychological constructs and employability?} Accounting for the significant demographic, behavioural, and geography biases, our analysis shows that employed males are more agreeable. Unemployed females exhibit high levels of openness. 
Moving to higher level constructs, we find that unemployed individuals tend to have a more social binding worldview, prioritising purity values. In particular, unemployed females are strongly valuing in-group loyalty and authority. 
Moral values reflect our political viewpoints and personal narratives.
Coherent with our previous findings, employed males are not really concerned with protecting their traditional values, while for them is fundamental the fair treatment of everyone in the society.

\textit{Are there divides between the generic interests and cultural attributes with respect to the occupational status?}
Self-reported scores did not show any striking differences to the general interests of employed and unemployed. Employed declared to be more interested in societal issues and hobbies, while unemployed in business and food pages.
The two communities have broadly the same interests; this finding of extreme importance since it shows that the employability status is not related with the interests or cultural differences but rather with the psychological aspects and life opportunities.

Assessing the linguistic content of the liked pages allows for a more in-depth understanding of peoples' interest. When comparing the linguistic content of pages likes by the two communities, the probability distributions (language models) show that the most distinguishing words for the unemployed are about hobbies and entertainment while for the employed about cultural projects and associations promoting the wine, food and in general the territory.
This finding leads us to the conclusion that the two communities may have similar interests but make different use of the platform.  
Excluding the commonly liked content, the NMF topic models confirm that the general interests of the two communities are along the same lines; however, in the prevalent topic of the unemployed community, news and media play a major role followed by sports. On the other side, the employed community focuses more on non-profit associations and personal blogs. This finding indicates that the unemployed population use Facebook as a source of information.
Such finding is alerting given the susceptibility of social media platforms to mis/disinformation ~\cite{cossard2020falling,kalimeri2019human,cinelli2020echo}. 

Concluding, our findings show small but significant differences in the psychological and moral values of the two communities. 
Importantly, it emerged that the two communities utilise the Facebook platform differently, with the unemployed to experience it as a news resource and the employed to connect with their local activities.
We believe that these findings can inform policymakers to devise better policies that do not solely improve the hard but also the soft skills of this fragile population.

%
%
%
\bibliographystyle{unsrt}  

\bibliography{UnemploymentTopic} 

\section{Appendix}

\vspace{-10pt}
\begin{figure*}[ht]
\centering
\includegraphics[width=.49\columnwidth,keepaspectratio]{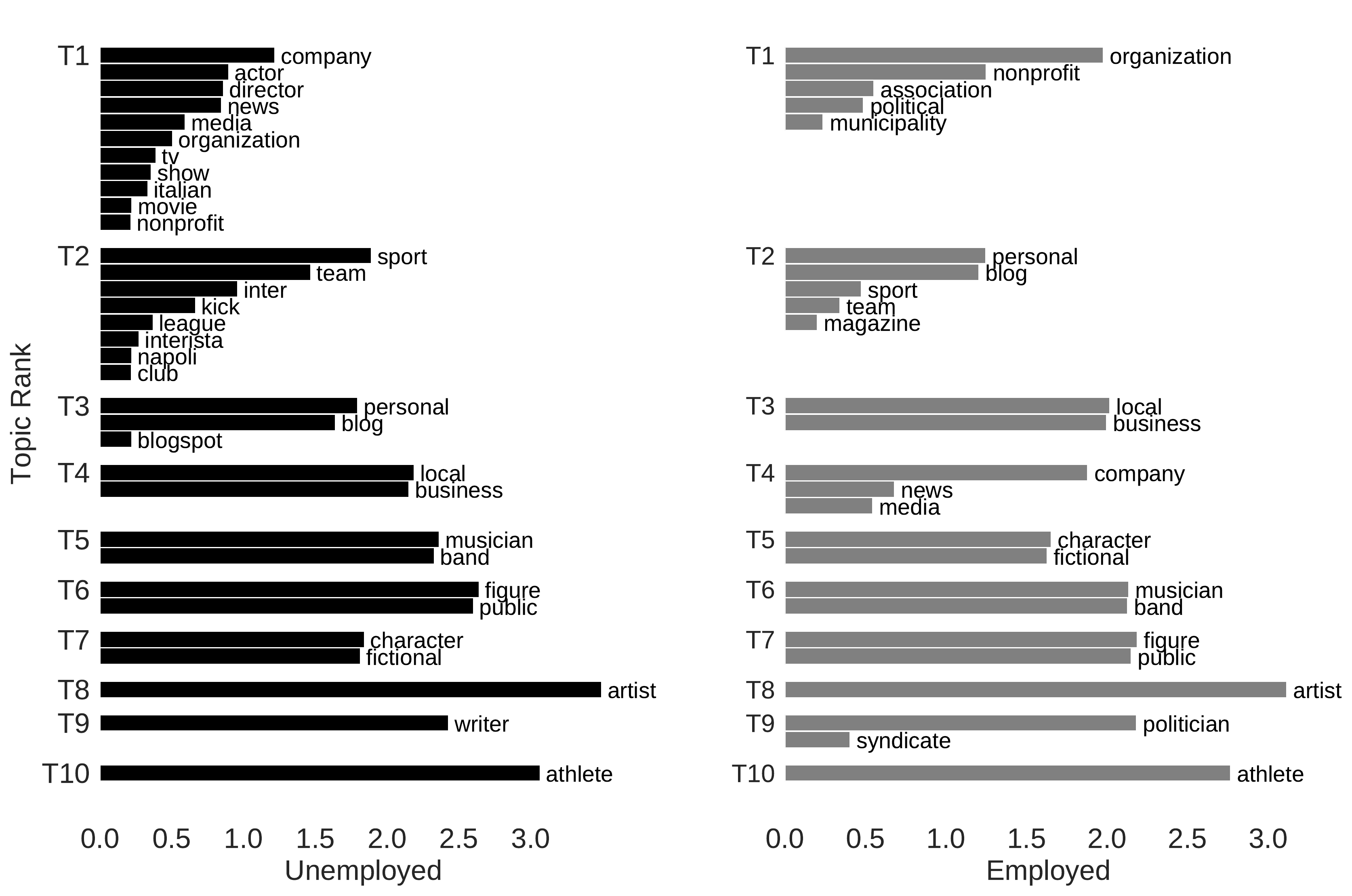}
\includegraphics[width=.49\columnwidth,keepaspectratio]{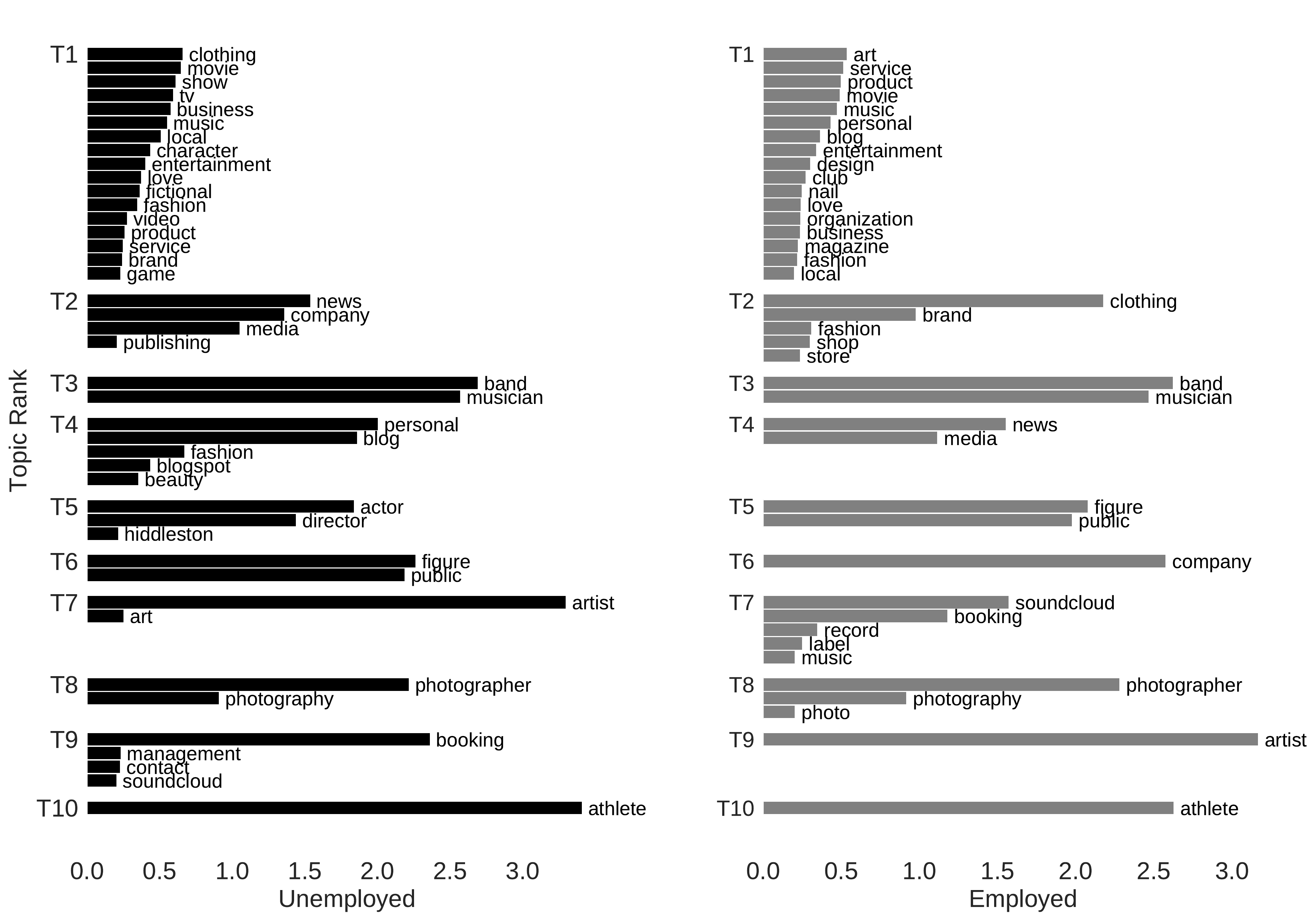}
\caption{The ten topics according to the NMF algorithm with the most contributing words per topic, for the Italian (first two columns) and the English language (last two columns). }
\label{fig:topic_italian}
\vspace{-10pt}
\end{figure*}

\begin{table*}[ht]
\tiny{
    \centering
    \renewcommand{\arraystretch}{1.2}

    \begin{tabular}{llllllll}
        \textbf{}&\textbf{D}&\textbf{MFT}&\textbf{D$+$MFT}&\textbf{B5}&\textbf{D$+$B5}&\textbf{PO}&\textbf{D$\pm$PO}\\
        \textit{n}&\textbf{ }3079&\textbf{ }786&\textbf{ }786&\textbf{ }687&\textbf{ }687&\textbf{ }1114&\textbf{ }1114\\
        $R_{MD}^{2}$&\textbf{ }0.075&\textbf{ }0.026&\textbf{ }0.09&\textbf{ }0.042&\textbf{ }0.084&\textbf{ }0.029&\textbf{ }0.050\\
        \toprule
        Intercept&-1.287(***)&-1.298&-1.123&-2.056(*)&-1.777&-1.857&-0.870\\\cmidrule{2-8} 
        Gender (M)&-1.007(***)&&-0.726(***)&&-0.686(***)&&-0.740(***)\\
        Center&\textbf{ }0.189(***)&&\textbf{ }0.080&&\textbf{ }0.168&&-0.104\\
        South&\textbf{ }1,052(***)&&\textbf{ }0.199(***)&&\textbf{ }1.217(***)&&\textbf{ }0.385\\\cmidrule{2-8}
        Care&&\textbf{ }0.052&\textbf{ }0.033&&\\
        Fairness&&-0.084(**)&-0.070&&\\
        Loyalty&&\textbf{ }0.010&\textbf{ }0.016&&\\
        Authority&&-0.008&-0.004&&\\
        Purity&&\textbf{ }0.066(**)&\textbf{ }0.046&&\\\cmidrule{2-8}
        Extraversion&&&&\textbf{ }0.002&-0.015\\
        Agreeableness&&&&\textbf{ }0.050&-0.060\\
        Conscientiousness&&&&\textbf{ }0.094&\textbf{ }0.103&\\
        Openness&&&&\textbf{ }0.056&\textbf{ }0.032&\\
        Neurotisism&&&&\textbf{ }0.011&\textbf{ }0.003\\
        \cmidrule{2-8}
        Q1&&&&&&\textbf{ }0.230&\textbf{ }0.244\\
        Q2&&&&&&-0.052&-0.044\\
        Q3&&&&&&-0.026&-0.254\\
        Q4&&&&&&-0.108&-0.132\\
        Q5&&&&&&\-0.080&-0.068\\
        Q6&&&&&&\textbf{ }0.071&\textbf{ }0.071\\
        Q7&&&&&&\textbf{ }0.019&\textbf{ }0.161\\
        Q8&&&&&&-0.050&-0.036\\
    \bottomrule
    \end{tabular}
    
    \caption{Logistic regression models predicting qualification using demographic (D), Big Five (B5), Moral Foundation (MFT), Political Opinion (PO). The Table shows alongside their coefficient estimate and their corresponding p-values (Bonferroni- adjusted). Confidence levels: p $<$ 0.001 ***, p $<$ 0.01 **, p $<$ 0.05 *.
    \vspace{-20pt}
\label{tab:regression}}}
\end{table*}

\end{document}